\documentclass[aps,prb,twocolumn,showpacs,floatfix]{revtex4}

\usepackage{graphicx}

\begin{document}

\title{Density functional study of the electronic and vibrational properties of TiOCl}

\author{L. Pisani$^1$, R. Valent\'\i$^2$, B. Montanari$^3$, and N. M. Harrison$^{1,4}$ }

\affiliation{$^1$Department of Chemistry, Imperial College London, South Kensington campus, 
 London SW7 2AZ, United Kingdom \\ $^2$Institut f\"ur Theoretische Physik, Universit\"at Frankfurt, 
60438 Frankfurt, Germany \\ $^3$STFC Rutherford Appleton Laboratory, Chilton, Didcot, 
Oxfordshire OX11 0QX, United Kingdom \\ $^4$STFC Daresbury Laboratory, Daresbury, Warrington WA4 4AD, 
United Kingdom}

\date{\today}

\begin{abstract}

\vspace{0.5cm}

\footnotesize{ 
We present the phonon spectrum of
TiOCl computed using hybrid density functional theory (DFT).
A complete analysis of the spectrum is performed for the space group $Pmmn$ (high symmetry phase) and
the space group $P2_1/m$ (low symmetry phase),  which is the symmetry of the spin-Peierls phase.
We show that the nonlocal correlations present in
the hybrid DFT approach are important for understanding the electron-lattice interactions in TiOCl.
The computed frequencies  compare well with those observed in Raman and infrared spectroscopy
experiments and we identify the origin of an anomalous phonon observed in Raman spectroscopy.
The relationship between relevant zone boundary
phonons in the high symmetry phase and the zone center counterparts in the $P2_1/m$ symmetry
allow us to speculate about the origin of the spin-Peierls phonon. }

\end{abstract}
\maketitle

\section{Introduction}
\label{intro}
The unconventional properties of the low-dimensional spin-1/2 compound TiOCl
have been a subject of intensive debate in recent 
years~\cite{Seidel_03,Lemmens_04,Saha_04,Lemmens_05,Rueckamp_05,Hoinkis_05,Pisani_05,shaz:100405,cv,Saha_05,Craco_06,dmft,Clancy_07,Hoinkis_07}.
This system shows two consecutive phase transitions at
$T_{c_2}$= 91K and at $T_{c_1}$= 60K from a magnetic
low-dimensional behavior to a spin-Peierls dimerized 
state\cite{Seidel_03,Lemmens_05} through an intermediate,
structurally incommensurate, phase.  While the conventional spin-Peierls  phase in a
spin chain is well understood in terms of the magnetoelastic
coupling, the behavior in TiOCl is more complicated. 
The $ab$-planes of TiOCl consist of TiO bilayers separated by layers of Cl$^-$ ions
stacked along the $c$ direction\cite{Seidel_03}. 
This bilayered structure, in conjunction with the localized nature of the 
Ti 3$d$ electrons, has motivated various proposals
about the role of interchain frustration~\cite{Rueckamp_05}
(within one layer and between layers), lattice degrees of freedom and correlation 
effects~\cite{Pisani_05,Hoinkis_05,Saha_05,dmft} at the phase transitions.
Initial suggestions about the influence of orbital 
fluctuations~\cite{Lemmens_04} have been settled both
experimentally and theoretically and the groundstate of the system 
at temperatures above $T_{c_2}$ is now known to be characterised  by singly occupied Ti 3$d_{xy}$ 
orbitals~\cite{Hoinkis_05,Rueckamp_05,Lemmens_05,Saha_05}. 

The ground state electronic structure of this system in the observed room
temperature structures has been calculated
within {\it ab initio} density functional theory (DFT)~\cite{Seidel_03,Saha_04,Lemmens_05,Hoinkis_05}.  
The role of electronic correlations beyond those included in the local density approximation (LDA)
have also been studied using the LDA+U~\cite{Seidel_03,Saha_04,Hoinkis_05},
LDA+DMFT~\cite{Saha_05,Craco_06}, and very recently
the LDA+cluster-DMFT\cite{dmft} approximations.
In the latter approach, where  two-site correlations
were considered, a good agreement between the calculated spectral function 
and the measured photoemission and O K-edge X-ray absorption spectra
was achieved; for all other methods mentioned the computed band gap is significantly below
that observed. These studies provide strong evidence for the important role played by
electron correlations in this material.

The interplay between the structural and
electronic degrees of freedom within the various phases,
which is particularly relevant in the study of the anomalous spin-Peierls
transition, has only been partially investigated from a theoretical
point of view. 
In a previous study by some of the authors~\cite{Pisani_05}, the Raman active
$A_g$ phonon modes in TiOCl for the high-symmetry (HS) phase ($T > T_{c_2}$)  
were obtained within an LDA+U formalism and the frozen-phonon approach.
It was concluded that  
the inclusion of electron correlation at the LDA+U level
significantly improved the agreement between the computed and observed frequencies.
In the present work we extend these studies to include the  $P2_1/m$ phase, which has been 
identified by Palatinus {\it et al.}\cite{Palatinus_05} as the low-symmetry (LS)
 phase ($T<T_{c_1}$).  We present the electronic structure
and a complete classification of the  phonon spectrum 
for both the HS and the LS structures, and make a detailed comparison
with the measured Raman and infra-red spectra.
We also extend the treatment of electron correlation to the hybrid exchange DFT 
approximation - which includes both the local (U) and non-local components of
exchange and correlation beyond those present in the LDA.  The much better
agreement with the observed spectra compared
to that achieved in previous calculations (within the LDA and LDA+U)~\cite{Pisani_05}
strongly suggests that nonlocal correlations are important
to describe the effects of lattice vibrations in TiOCl. 
Within this theoretical framework we
can also identify for the first time the anomalous phonon observed in Raman 
spectroscopy~\cite{Lemmens_04} which appears to violate the Raman selection rules. 
Moreover,  by analysing the relationship  between
the zone boundary phonons in the $b$ direction (Ti-chain axis)
in the HS phase and relevant zone center phonons in the $P2_1/m$ phase,
we suggest possible candidates for the spin-Peierls phonon mode.

The paper is organized  as follows: in section II we
present the computational details of our calculations, in section
III  we discuss  the electronic structure of TiOCl in the 
HS phase within the various approximations used in previous studies
and present our results using  hybrid exchange DFT.
Sections IV and V report on the optimization of
the lattice parameters and the calculation of phonons for the HS phase.
In section VI we compute and discuss the phonons
in the  $P2_1/m$ symmetry (LS phase) and in the last section we summarize 
our conclusions.

\section{Computational Details}

The first-principles calculations presented here have been
performed using the hybrid exchange density functional
B3LYP~\cite{b3lyp1,b3lyp2} as implemented in the CRYSTAL package~\cite{CRYSTAL06}. 
In CRYSTAL, the crystalline wave functions are expanded as a linear
combination of atom centered Gaussian orbitals LCAO
with $s$, $p$, $d$, or $f$ symmetry. 
The Ti, O and Cl atoms are described by a triple
valence all-electron basis sets: an 86-411G** contraction (one $s$,
four $sp$, and two $d$ shells), an 8-411G* contraction (one
$s$, three $sp$, and one $d$ shells) and an 86-311G* (one
$s$, four $sp$, and one $d$ shells),  respectively.
A reciprocal space sampling on a
Monkhorst-Pack grid\cite{Monkhorst_76} of shrinking factor equal to 6 is
adopted after finding it to be sufficient to converge the total
energy to within 10$^{-6}$ eV per unit cell. The Gaussian overlap
criteria which controls the truncation of the Coulomb and
exchange series in direct space are set to 10$^{-7}$, 10$^{-7}$, 10$^{-7}$,
10$^{-7}$, and 10$^{-14}$ the details of which are presented elsewhere~\cite{CRYSTAL06}. 
Typically linear mixing of 70\% and an
Anderson second-order mixing scheme was used to guide the
convergence of the self-consistent field procedure. 

The geometry optimizations are performed using the algorithm
proposed by Schlegel {\it et al.}~\cite{schlegel}.
The full dynamical matrix was computed using finite differences of the
analytic gradients for which the atoms were displaced by 0.02 \AA.

\section{Electronic structure}
\label{elstr}

In Ref.~\onlinecite{Pisani_05}, the authors performed an investigation 
of the electronic  structure of TiOCl at the experimental cell parameters of 
the HS phase (space group $59$  $Pmmn$)  using the
LDA and LDA+U methods. 
It was concluded that the LDA functional
does not reproduce the measured electronic structure or the internal coordinates
of the atoms in the cell, 
whereas within the LDA+U method the  Mott-Hubbard insulator nature
of TiOCl is obtained as well as the internal coordinates.
Nevertheless, the LDA+U approximation does not
describe properly the optical band-gap of 2 eV observed
in infrared spectroscopy experiments. While the gap between the occupied
spin up $d_{xy}$ band and the unoccupied spin down $d_{xy}$ is
computed to be  2.8 eV within the LDA+U model using U and J values of 3.3 eV 
and 1 eV respectively, the gap between the occupied $d_{xy}$
and unoccupied $d_{xz-yz}$ of the same spin is only 0.3 eV.
Since the optical gap is a measure of the lowest energy difference between
occupied and unoccupied electronic states, the LDA+U calculations significantly
underestimate the measured value.

Saha-Dasgupta {\it et al.} improved on the description of on-site
correlation by using a single-site LDA+DMFT, where dynamical fluctuations
beyond the LDA+U are introduced using a quantum monte carlo
 impurity solver~\cite{Saha_05}. 
This method provides a better
description of the overall width of the spectral function,
but the calculated band gap of 0.5 eV still underestimates very
significantly the observed gap. 
LDA+DMFT calculations performed with an iterated
perturbation theory (ITP) impurity solver~\cite{Craco_06}
yield a bigger energy gap at the cost of introducing a mixing of the $d$-orbitals
significantly greater than that deduced from the spectroscopy~\cite{Rueckamp_05,Hoinkis_05}.
 
Based on the fact that TiOCl is a low-dimensional
correlated system and that at low temperature it
dimerizes into Ti-Ti singlet 
pairs along the $b$-axis, it has been suggested~\cite{dmft} 
that two-site correlations may be important even at high temperatures and thus that the
single site approximation will be insufficient.
We note  that in Raman spectroscopy~\cite{Lemmens_04}, specific heat~\cite{cv} and
NMR experiments~\cite{nmr} strong  evidence is found 
for the existence of low energy fluctuations
of the spin system in a temperature range that extends well above the 
spin-Peierls transition. 
The  role of inter-site  correlation
has been investigated previously in Ref.~\onlinecite{dmft} by using the cluster extension
of DMFT in which the correlated impurity is taken to be Ti-Ti pairs instead of a single site~\cite{Saha_05}.
The computed linewidth, the lineshape and the band gap of the calculated spectral function
are found to be in good agreement with angle integrated photoemission data and with
O K-edge absorption spectra\cite{dmft}.  

In the following,  we start from a completely different approach
and demonstrate that the role of nonlocal correlations can be also
captured in an effective manner within the hybrid DFT approach.

The LDA+U scheme is known to lack any description of non-local (off-site) 
electron exchange and correlation beyond the local density approximation (LDA)
due to the fact that the empirical Hubbard U term added to the Hamiltonian improves only
on the description of on-site Coulomb effects.
In order to capture the non-local component of the electronic exchange and correlation  
(which is unsatisfactorily approximated within the LDA) one would need to extend 
the on-site LDA+U treatment to inter-site interactions, which therefore would take 
the form of an extended version of the LDA+U method, namely the LDA+U+V scheme.
Within this scheme, the inter-site term V is treated within the Hartree-Fock (mean-field) approximation
(as the U term is in LDA+U) and thus the scheme is sufficient to introduce the
Ti-Ti pair interaction neglected in the LDA.
In hybrid DFT, the treatment of off-site exchange and correlation effects beyond LDA
is already built in due to the inclusion in the Hamiltonian of a Fock-exchange term  
which operates between all (on-site and off-site) Kohn-Sham orbitals. The effectiveness
of this term for describing the on-site correlation is well documented~\cite{feng}
and there is every reason to believe that it will be equally effective in describing the non-local exchange
and correlation effects beyond LDA at a mean-field level.   

Fig.~\ref{band} shows the spin polarized band structure of TiOCl 
calculated within the B3LYP approximation for the experimental HS structure~\cite{shaz:100405}.  
We note that the band structure does not
change significantly when calculated at the relaxed HS geometry.
We observe that, in contrast with the LDA+U approach, the adoption of the  B3LYP approximation improves 
significantly on the description of the electronic ground state by opening a band gap 
of about 2.2 eV, in good agreement with the measured gap.
The band states just below the Fermi level are majority spin states of $d_{xy}$
character (Fig. \ref{band}, left panel). The coordinate system has
been chosen with $z$ in the same direction as $a$,
and with the $x$ and $y$ axes rotated by $45^o$ with respect to $b$ and $c$.
Above the Fermi level on the majority spin channel (Fig. \ref{band}, 
left panel) we find, in order of increasing energy,
band states with $d_{xz-yz},d_{xz+yz},d_{x^2-y^2},d_{z^2}$ character and 
likewise in the minority spin panel (Fig. \ref{band},  right panel), with
the only difference  that the $d_{xy}$ spin down states are unoccupied 
and placed between the $d_{xz-yz}$ and the $d_{xz+yz}$ states.
It is worth pointing out that the bands in Fig.~\ref{band} 
have been computed for a ferromagnetic alignment of Ti spins.
In the antiferromagnetic case  (not shown) 
the band structure of up and down states  is symmetric, 
as in any conventional antiferromagnet, and
the band gap is found to be the same as in the ferromagnetic case.
The states at the boundary of the 
band gap are  of $d_{xy}$ symmetry (occupied) and of $d_{xz-yz}$ symmetry (unoccupied).

The above features on the orbital nature of the states around the band gap are
 in full agreement with previous results obtained
 within LDA+U~\cite{Seidel_03,Saha_04}.

\begin{center}
\vspace{.2cm}
\begin{figure}[!h]
\includegraphics[scale=.35,clip]{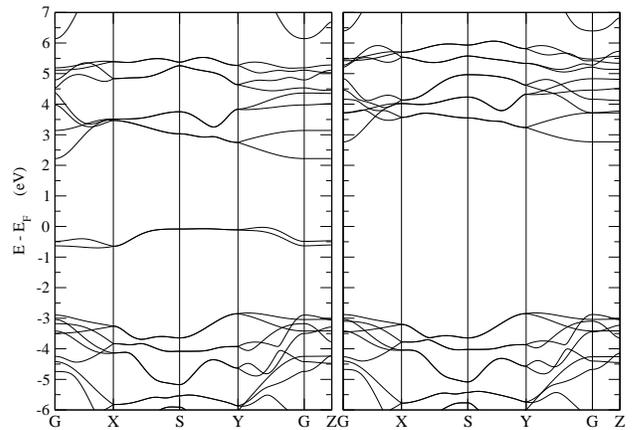}
\caption{Calculated band structure for the high symmetry phase of TiOCl 
  at the experimental crystal structure along the path $\Gamma$-X-S-Y-$\Gamma$-Z. 
  ($\Gamma$[0,0,0]; X[1/2,0,0]; S[1/2,1/2,0]; Y[0,1.2,0]; Z[0,0,1] )
  The majority and minority spin bands are shown in the
  left and right panel, respectively. $E_F$ indicates the Fermi energy.  }
\label{band}
\end{figure}
\end{center}

\section{Lattice Structure}
\label{vol}

In Table~\ref{abc}  we present the equilibrium lattice
parameters of the HS phase obtained after a full relaxation 
of the atomic structure of the FM state within  the B3LYP approximation.
For comparison,  we also show the results obtained with the
gradient corrected functional of Perdew, Burke and Ernzerhof~\cite{pbe} (PBE)
and the experimental values\cite{shaz:100405}.

While the agreement between the calculated and measured $a$  parameter 
is reasonably good,  this is not the case for $c$ and $b$.
The strong disagreement in the case of the $c$ parameter
is due to the predominantly van der Waals type of binding 
between bilayers in the $c$ direction
which is not correctly described in LDA, GGA or B3LYP approximations to DFT.
Keeping $c$ fixed at the experimental value and fully optimising the structure
does not change significantly the values of the $a$ and $b$ parameters.
We notice that the energy landscape is rather flat when varying the length of the $c$
lattice constant due to the very weak binding. However, we find no significant effect of 
the $c$ parameter on the assignment of the phonon modes in the Raman and
infrared spectra in the range of frequencies considered in the next section.
   
Concerning the parameter $b$ (chain direction), the B3LYP functional
gives an   overestimation of $b$ of about 6 \%.  
In order to understand this issue, we employed first the PBE functional 
for the full structural optimisation of the HS phase and then
compared with B3LYP results. 
With the choice of the PBE functional, we find that the structure is unstable with 
respect to Ti interchain dimerisation, where neighbouring chains shift rigidly towards each other,
leaving the intra-chain step unchanged and producing an alternating
interchain Ti-Ti distance.   
This is due to the fact that the electronic structure is found to be
a fully spin polarised  metallic state, due to the self-interaction error (SIE)
present in the GGA. The orbital polarisation found in SIE compensated functionals (LDA+U and B3LYP)
is lost within PBE and the orbital $d_{xy}$ becomes partially unoccupied in favour of the
$d_{xz-yz}$ orbital.  The latter  points towards Ti atoms of  neighbouring chains
and therefore favours interchain dimerisation.

This observation is useful in interpreting the behaviour of the B3LYP hybrid functional.
We note here that the functional of Lee, Yang and Parr\cite{lyp} (LYP) 
that describes the exchange-correlation term in B3LYP, 
belongs to the family of the PBE functional,
and thus we expect it to provide similar results as PBE. 
However, the inclusion of Fock exchange removes the interchain instability
found within the PBE functional and enables the B3LYP 
approximation to describe reasonably well the geometric and electronic structure
simultaneously. 

By changing the amount of Fock exchange from the amount prescribed for the B3LYP functional,
we are able to conclude that the overestimation of $b$ is strongly related to the percentage of the Fock exchange term
in the Hamiltonian.
Possible explanations of the disagreement between 
 the experimental and theoretical  values of $b$ are the following: 
 (i) the exchange contribution in B3LYP is overestimated due to the neglect of 
screening leading to an overestimation of the local Pauli pressure
and thus a stretching of $b$. 
 (ii) more speculatively, if 
 dynamical spin singlet fluctuations are important, they
could bring 
the Ti atoms into a precursor dimerised state with no long-ranged crystallographic order.
As a result, the x-ray measurement would probe a thermally averaged value of the Ti-Ti distance 
(=b/2) that is shorter than the ground state value.
It is not clear at present which of these explanations accounts for the overestimation of the $b$ parameter.

In Table~\ref{dist} we show the relevant equilibrium atomic distances and angles. 
In general a very good agreement of B3LYP results with the experimental values is obtained.

\begin{table}[!h]
\caption{Equilibrium lattice parameters of the high
 symmetry unit cell.}
\begin{tabular}{cccc}
\hline
\hline
      &  a (\AA) &  b (\AA)  &  c (\AA)   \\
\hline
 exp.  &  3.78 &  3.34 &  8.03  \\
 PBE   &  3.98 &  3.31 &  8.57  \\
 B3LYP &  3.81 &  3.49 &  8.69  \\
\hline
\label{abc}
\end{tabular}
\end{table}

\begin{table}[!h]
\caption{Equilibrium atomic distances and angles in the high symmetry unit cell.}
\begin{tabular}{ccccc}
\hline
\hline
       & Ti-O$|_{a}$ (\AA) &  Ti-O$|_{b}$ (\AA)  &  Ti-Cl (\AA) & Ti-O-Ti$|_{a}$   \\
\hline
 exp.  &    1.97     &       2.18    &    2.41  &  147.7$^\circ$     \\
 PBE   &    2.02     &       2.13    &    2.43  &  160.9$^\circ$     \\
 B3LYP &    1.98     &       2.22    &    2.45  &  148.6$^\circ$     \\
\hline
\label{dist}
\end{tabular}
\end{table}

\section{Phonons  in the pace group $Pmmn$}
\label{ht}

Within space group 59 ($Pmmn$), identified as the high 
symmetry phase of TiOCl, the $\Gamma$ point phonons are classified according to symmetry as follows:
$$
\Gamma=3A_g+3B_{2g}+3B_{3g}+2B_{1u}+2B_{2u}+2B_{3u}
$$
Each mode involves atomic displacements along a single axis. \\
The modes with symmetry $A_g$, $B_{2g}$, and $B_{3g}$ are along the $c$ , $a$
and $b$ axis, respectively, and are Raman active.
The modes with symmetry $B_{1u}$, $B_{2u}$, and $B_{3u}$ are along the $c$,
$b$ and $a$ axis, respectively, and composed  of  one acoustic and two optic modes,
the latter being infrared active.\\
In Table~\ref{HTph} the calculated frequencies
are reported and classified according to direction of displacement and irreducible
representation.

\begin{table}
\caption{ Phonon frequencies in the high symmetry phase of TiOCl.
The first and second column denote respectively  the irreducible representation
and the multiplicity  of the modes. The third column
indicates the direction of the phonon displacements. The phonon
frequencies are given in the fourth 3-set columns and we include
the reported experimentally observed frequencies in the fifth 3-set
columns. }

\begin{tabular}{cccccc|ccc}
          \multicolumn{9}{c}{RAMAN} \\
\hline
\hline
    Ir.Rep.  & Mult. & Displ.     &  \multicolumn{3}{c}{Freq.} &  \multicolumn{3}{c}{(experim.)}  \\
\hline
    $A_g$    & 3    & $c$       &  205  & 342   & 443  &  (203)  &  (365)  & (430)     \\
   $B_{2g}$ &  3    & $a$       &  116  & 256   & 613  &         &         &           \\
   $B_{3g}$ &  3    & $b$       &  141  & 287   & 431  &         &         &           \\
\hline
\end{tabular}

\vspace{0.5cm}

\begin{tabular}{cccccc|ccc}
            \multicolumn{9}{c}{INFRARED} \\
\hline
\hline
    Ir.Rep.  & Mult. & Displ.     &  \multicolumn{3}{c}{Freq.} &  \multicolumn{3}{c}{(experim.)}  \\
\hline
   $B_{1u}$ &  3    & $c$       &   ---   & 328   & 514 &       &         &        \\
   $B_{2u}$ &  3    & $b$       &   ---   & 256   & 274 &   ---   & (177)   & (294)  \\
   $B_{3u}$ &  3    & $a$       &   ---   & 154   & 463 &   ---   & (104)   & (438) \\
\hline
\end{tabular}

\label{HTph}

\end{table}

The frequencies of the $A_g$ modes at 205, 342 and 443 cm$^{-1}$ compare quite
well with the experimental values~\cite{Lemmens_04} as do   
the values computed within the LDA+U method in Ref.~\onlinecite{Pisani_05}.
No cross polarisation of light was considered in Ref.~\onlinecite{Lemmens_04} and therefore
no experimental values for $B_{2g}$ and $B_{3g}$ are available. 

Comparison with infrared measurements~\cite{Caimi_04} for the
$B_{2u}$ and $B_{3u}$ modes, is quite satisfactory
for the high frequency modes but less so for the low frequency ones.
As a possible explanation for the discrepancy, we estimated the anharmonic corrections
to the low frequency modes  by examining the energy surface 
along the mode and found it to be harmonic to an 
excellent approximation.
However, it is worth mentioning  that the infrared active modes  
have been measured separately  by Grueninger {\em et al.}~\cite{gruen} 
and a value of 201 cm$^{-1}$ was found for the lower frequency $B_{2u}$ mode,
which agrees significantly better with the calculated frequency. 

\section{Phonons in the space group $P2_1/m$}

\subsection{Phonon spectrum}

Within space group 11 ($P2_1/m$), which has been identified
by Palatinus {\it et al.}\cite{Palatinus_05} as the spin-Peierls
(LS) symmetry of TiOCl,
the $\Gamma$ phonons are classified as follows: 
$$
\Gamma=12A_g+10B_u+6B_g+5A_u
$$
where one $A_u$ and two $B_u$ modes are acoustic.

\begin{table}
\caption{Phonon frequencies in the low symmetry phase. The first four
columns have the same assignment as in Table \protect\ref{HTph}.
In the fifth column we include, where appropriate, the HS irreducible representation counterpart.}
\begin{tabular}{cccccccccc}
      \multicolumn{10}{c}{RAMAN}\\
\hline
\hline
Ir.Rep. & Mult. & Displ. & \multicolumn{6}{c}{Freq.} & SG59\\
\hline
$A_g$ &  3  &  $c$  &  203 & 342  &  443 &     &      &      &  $A_g$    \\
      &  3  &  $b$  &  140 & 289  &  431 &     &      &      &  $B_{3g}$ \\
      &  6  &  $bc$ &  175 & 212  &  285 & 323 & 382  &  490 &           \\
$B_g$ &  3  &  $a$  &  118 & 257  &  616 &     &      &      &  $B_{2g}$ \\
      &  3  &  $a$  &  122 & 199  &  540 &     &      &      &           \\
\hline
\end{tabular}

\vspace{0.5cm}

\begin{tabular}{cccccccccc}
      \multicolumn{10}{c}{INFRARED} \\
\hline
\hline
Ir.Rep. & Mult. & Displ. & \multicolumn{6}{c}{Freq.} & SG59\\
\hline
$B_u$ &  3 &  $c$  &  --- & 329  &  516 &     &      &      &  $B_{1u}$    \\
      &  3 &  $b$  &  --- & 257  &  276 &     &      &      &  $B_{2u}$    \\
      &  6 &  $bc$ &  175 & 212  &  285 & 323 & 382  &  490 &              \\
$A_u$ &  3 &  $a$  &  --- & 159  &  464 &     &      &      &  $B_{3u}$    \\
      &  3 &  $a$  &  122 & 199  &  540 &     &      &      &               \\
\hline 
\end{tabular}

\label{ltfreq}

\end{table}

 In this space group, the  A$_g$ and B$_g$ modes are Raman active
 while the A$_u$ and B$_u$ modes  are infrared active.
As a general feature, modes in this LS phase are either displacements within the  $bc$ plane
or purely along the $a$ axis. In particular, the A$_u$ and B$_g$ modes are along the $a$ axis
and the A$_g$ and B$_u$ are within the $bc$ plane.

We note that starting from the experimental
LS structure~\cite{Palatinus_05} within the space group 11 
and fully optimising the structure
results in a structure very close to the HS one with a unit cell
doubled along the $b$-axis. This implies that the optimised
structure does not show any signature of a spin-Peierls distortion. 
This is due to the fact that in the spin-Peierls mechanism
the loss of elastic energy is overcompensated
by an energy gain in spin-singlet fluctuations. 
The latter are not properly described within   
the current formulation of spin-DFT and therefore no magnetoelastic 
structural distortion can be expected for any of the exchange-correlation
approximations currently in use.
A consequence of this is that our
zone center phonon mode calculations for the LS phase  provide
a  good description of some relevant zone boundary phonons of the high symmetry
phase, namely those phonons that are associated with a doubling
of the HS unit cell along the $b$ axis and which are directly related
to the spin-Peierls instability.

In Table~\ref{ltfreq} the calculated frequencies and the directions of the corresponding displacement 
eigenvectors (within the reference system $a,b,c$ of the orthorhombic unit cell) 
are shown. Due to the absence of distortional features in the
structurally optimised LS structure, the 18 modes of the HS phase 
are seen to evolve unchanged into the LS phase and the additional 18 modes, 
originating from the doubling of
the unit cell, have degenerate features which will be discussed below.

In Table~\ref{ltfreq} we choose to decompose the LS modes according to their HS irreducible 
representation counterparts, where meaningful.
It is important to note, however, that the spin-Peierls distortion does not modify  
the Raman active A$_g$ modes' frequencies of the HS phase which are seen to evolve 
unchanged in the LS Raman spectra~\cite{Lemmens_04}. 
We can therefore assume that the effect of lattice distortion
is not crucial as far as the frequency calculation is concerned.

\begin{figure*}

\begin{center}
\mbox{A$_g$}
\begin{minipage}{5.5cm}
175 cm$^{-1}$
\includegraphics[width=5.5cm]{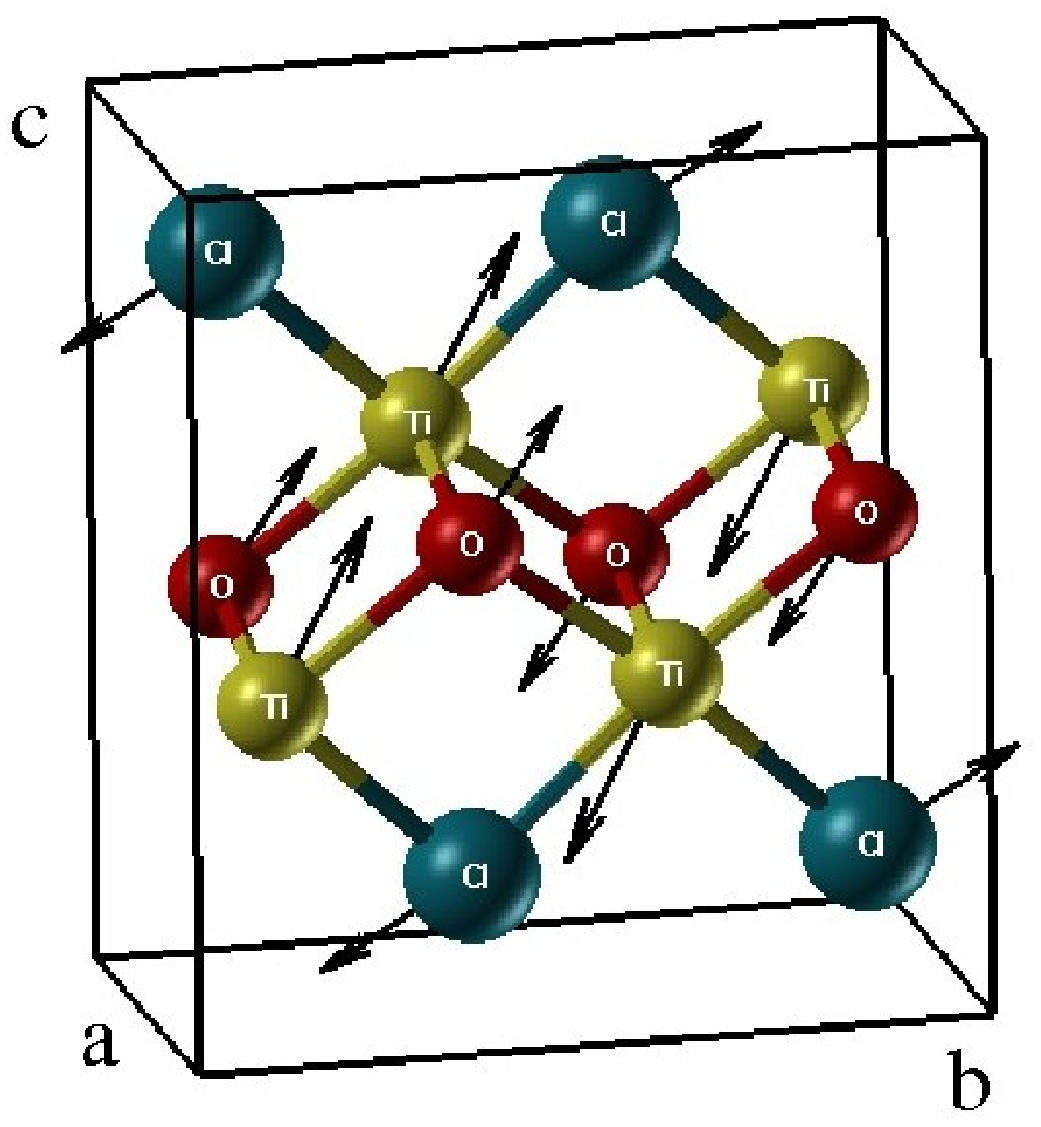}
\end{minipage}
\begin{minipage}{5.5cm}
212 cm$^{-1}$
\includegraphics[width=5.5cm]{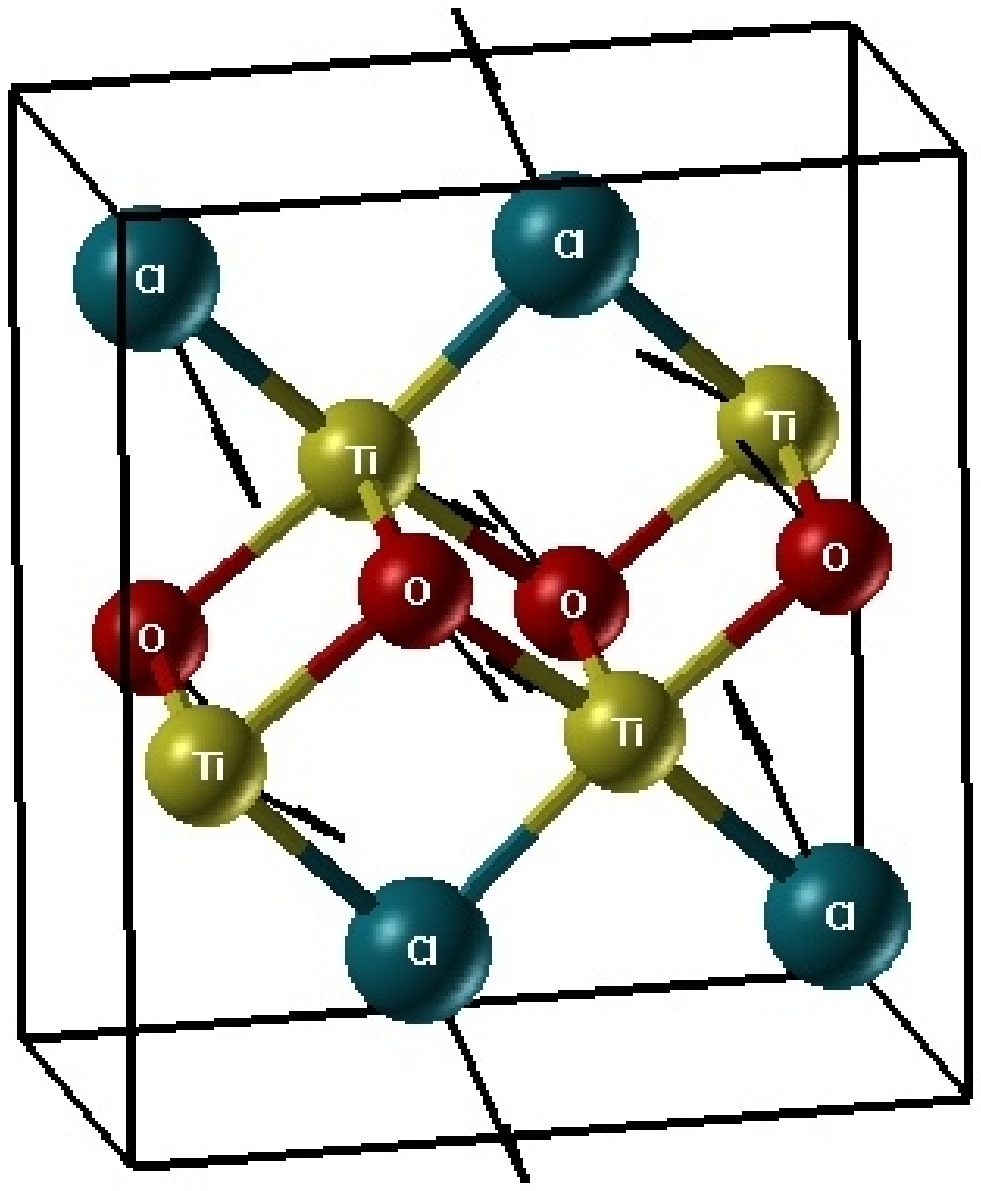}
\end{minipage}
\begin{minipage}{5.5cm}
285 cm$^{-1}$
\includegraphics[width=5.5cm]{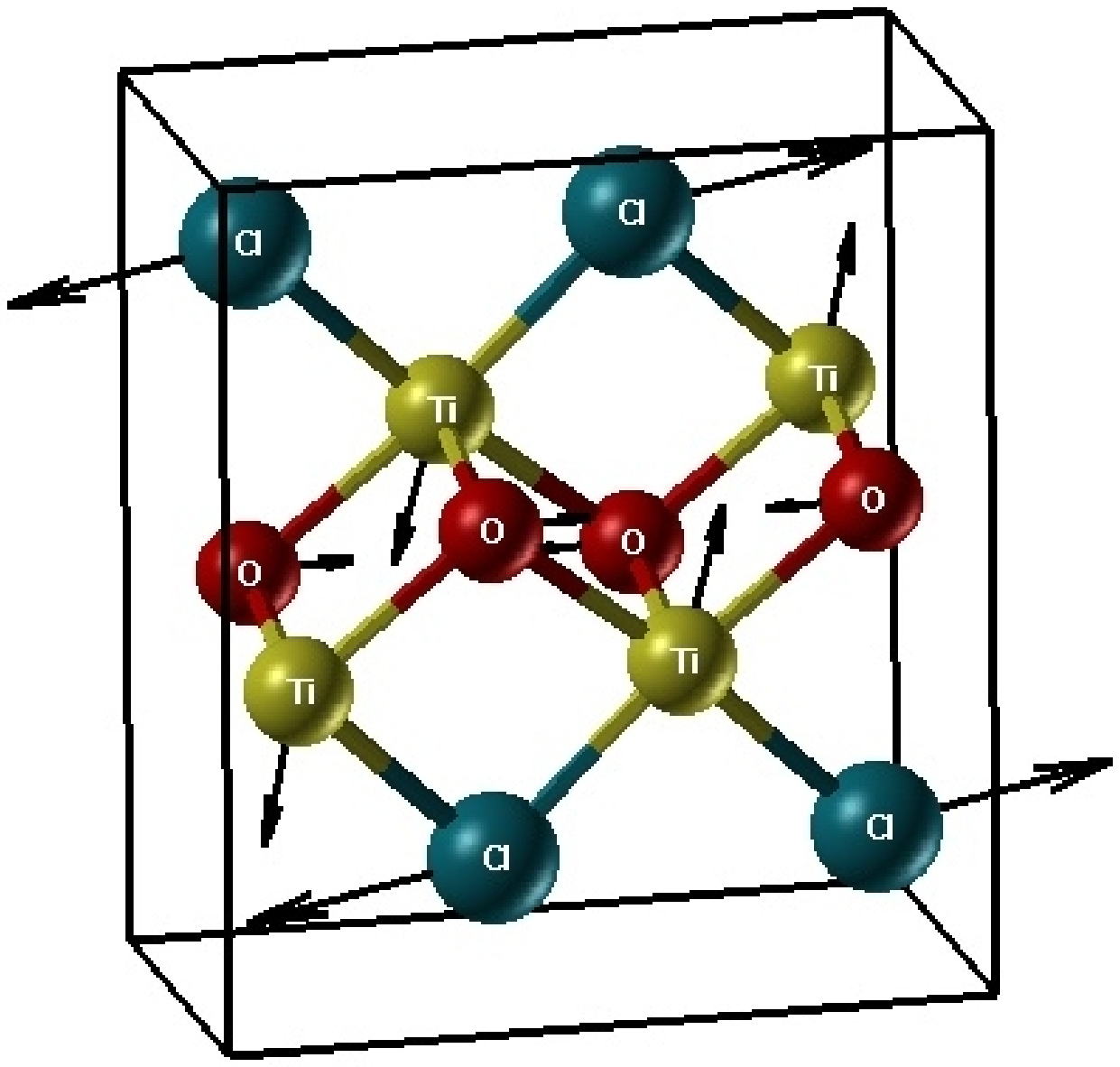}
\end{minipage}
\end{center}
\vspace{-1cm}
\begin{center}
\mbox{B$_u$}
\begin{minipage}{5.5cm}
\includegraphics[width=5.5cm]{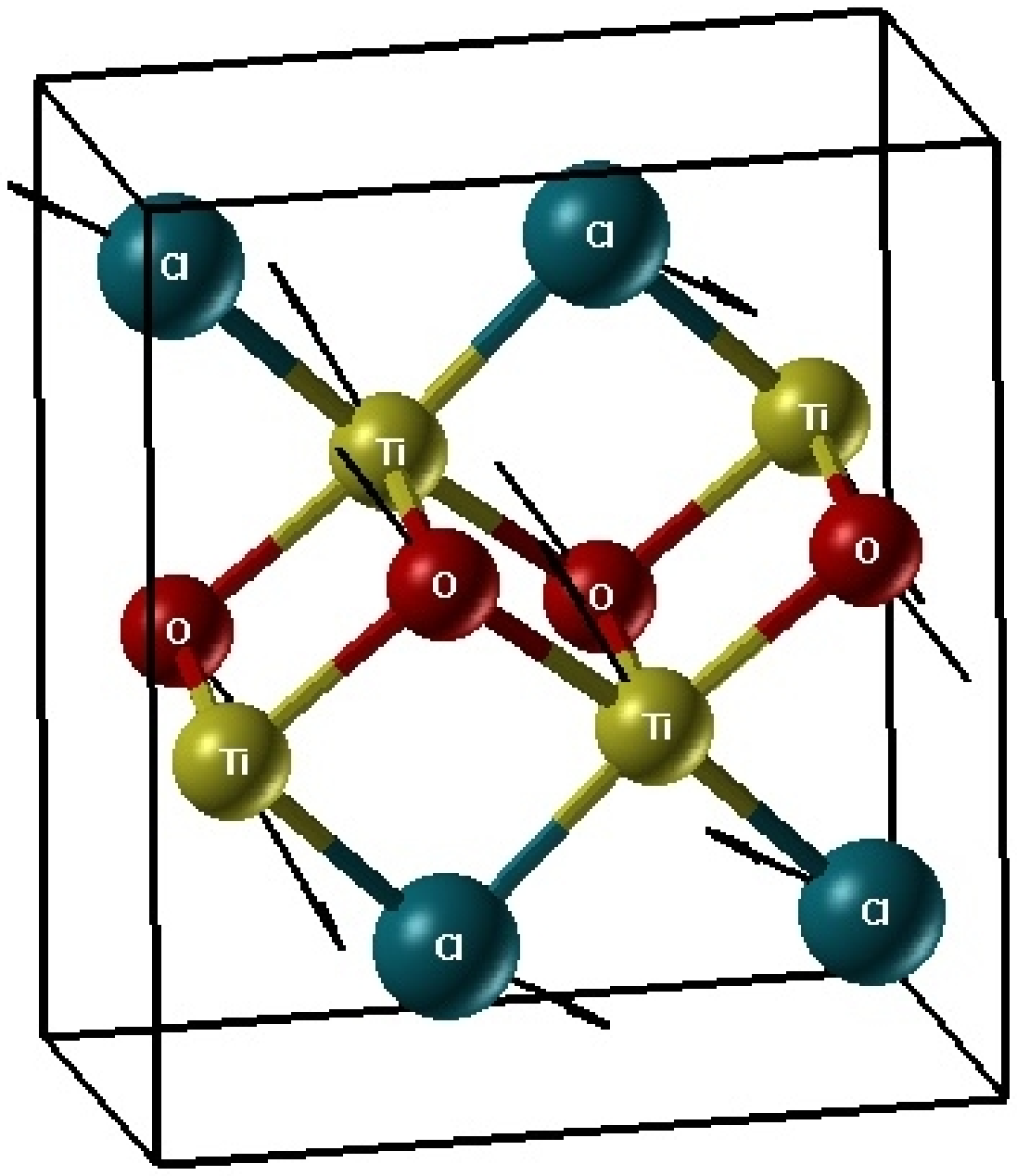}
\end{minipage}
\begin{minipage}{5.5cm}
\includegraphics[width=5.5cm]{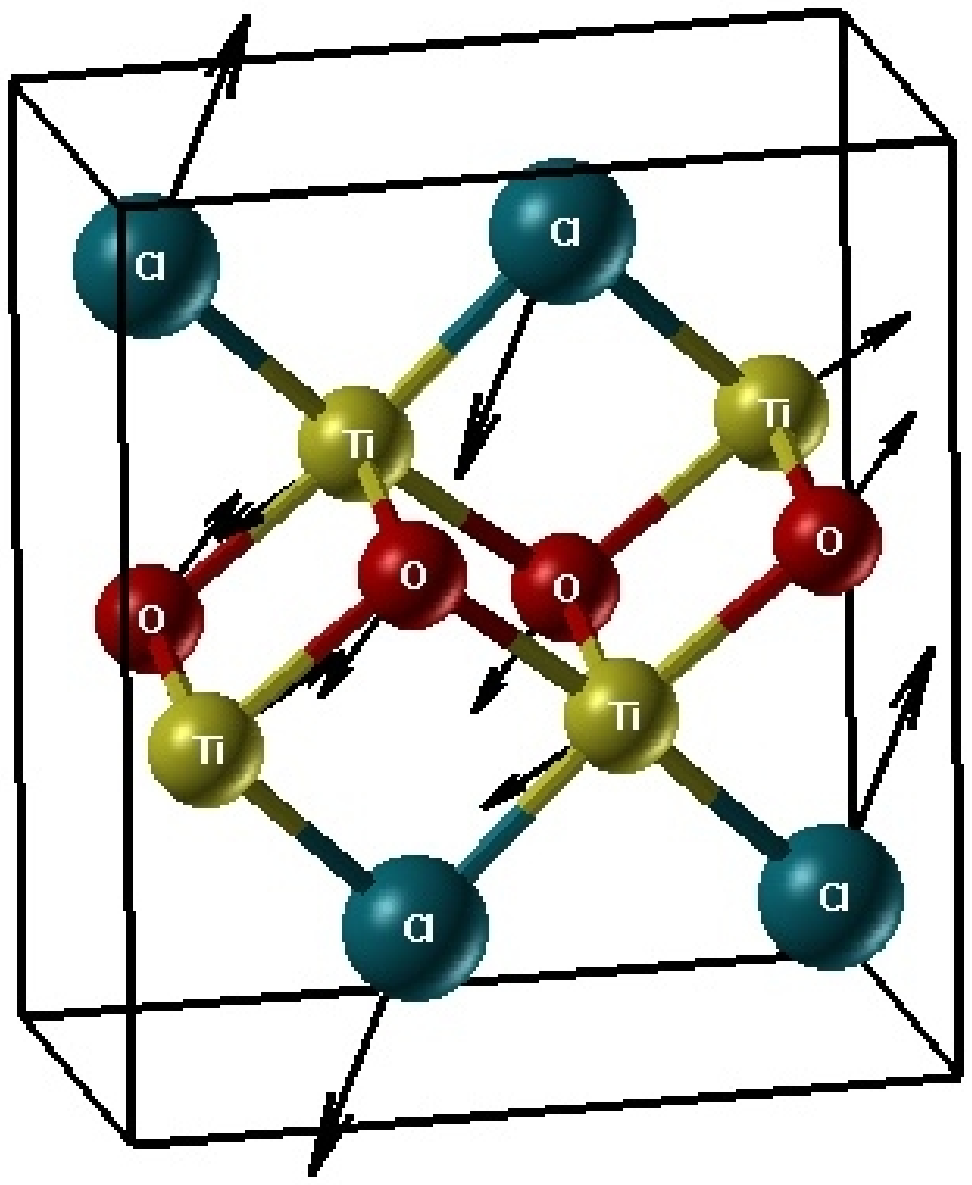}
\end{minipage}
\begin{minipage}{5.5cm}
\includegraphics[width=5.5cm]{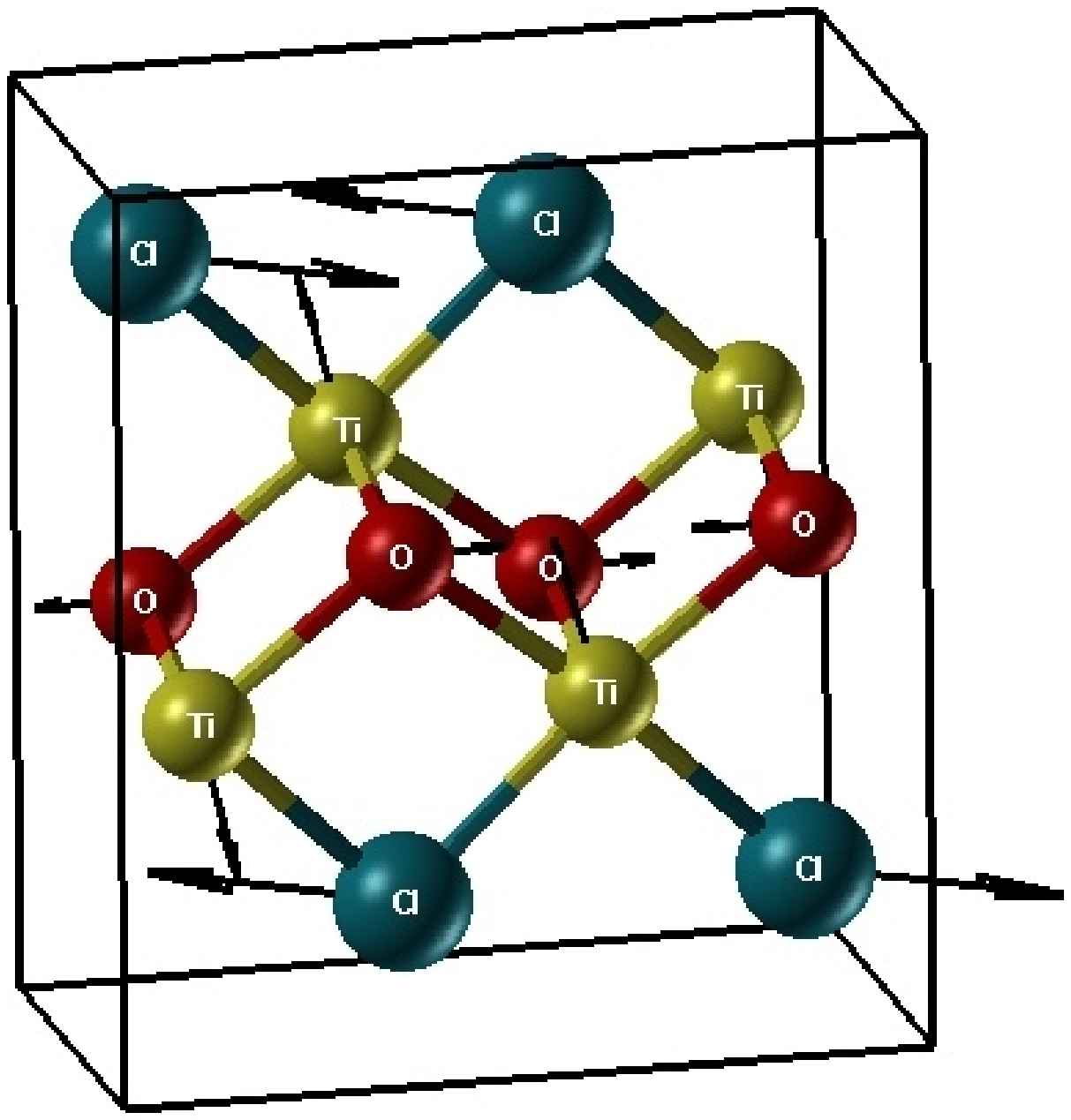}
\end{minipage}
\end{center}

\vspace{1cm}

\begin{center}
\mbox{A$_g$}
\begin{minipage}{5.5cm}
323 cm$^{-1}$
\includegraphics[width=5.5cm]{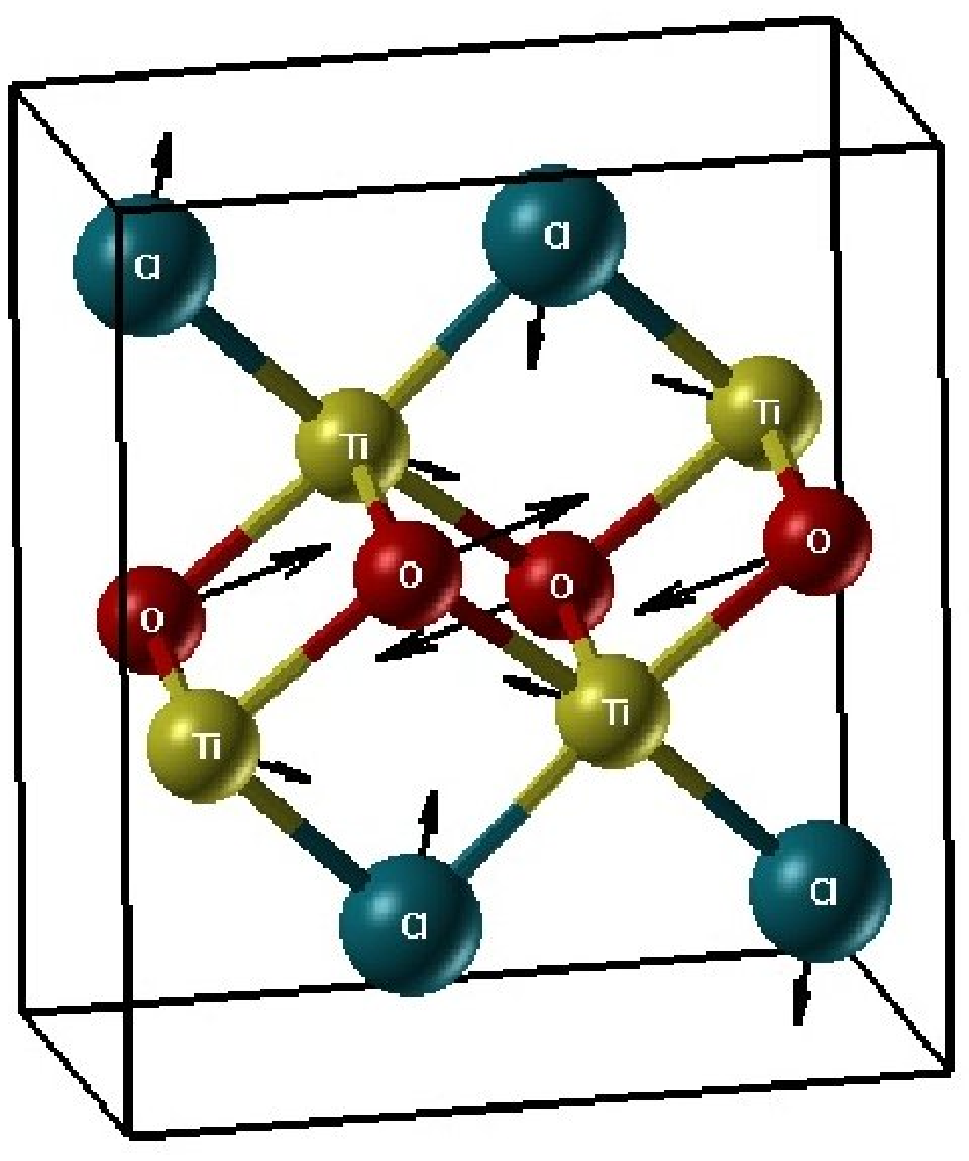}
\end{minipage}
\begin{minipage}{5.5cm}
382 cm$^{-1}$
\includegraphics[width=5.5cm]{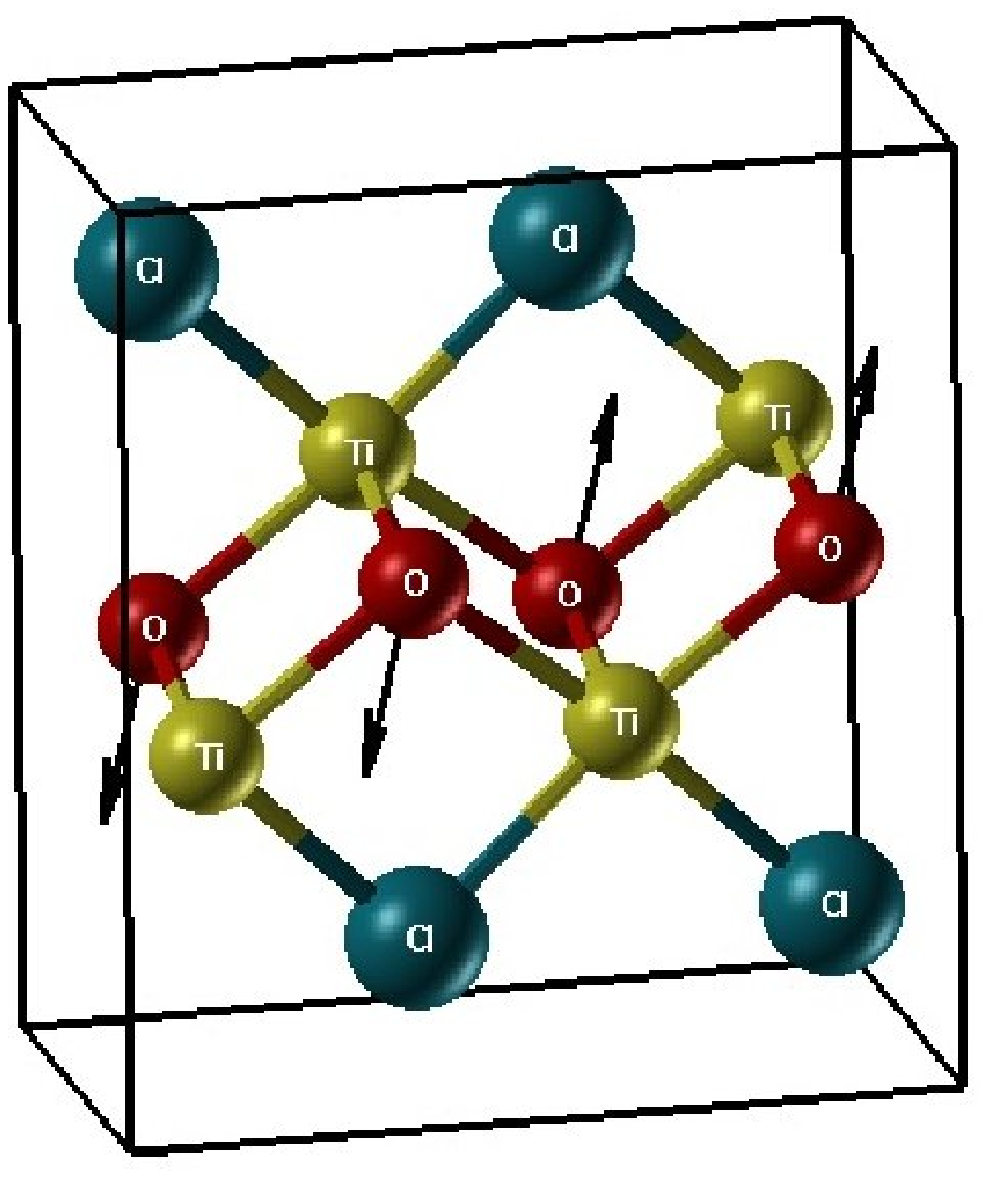}
\end{minipage}
\begin{minipage}{5.5cm}
490 cm$^{-1}$
\includegraphics[width=5.5cm]{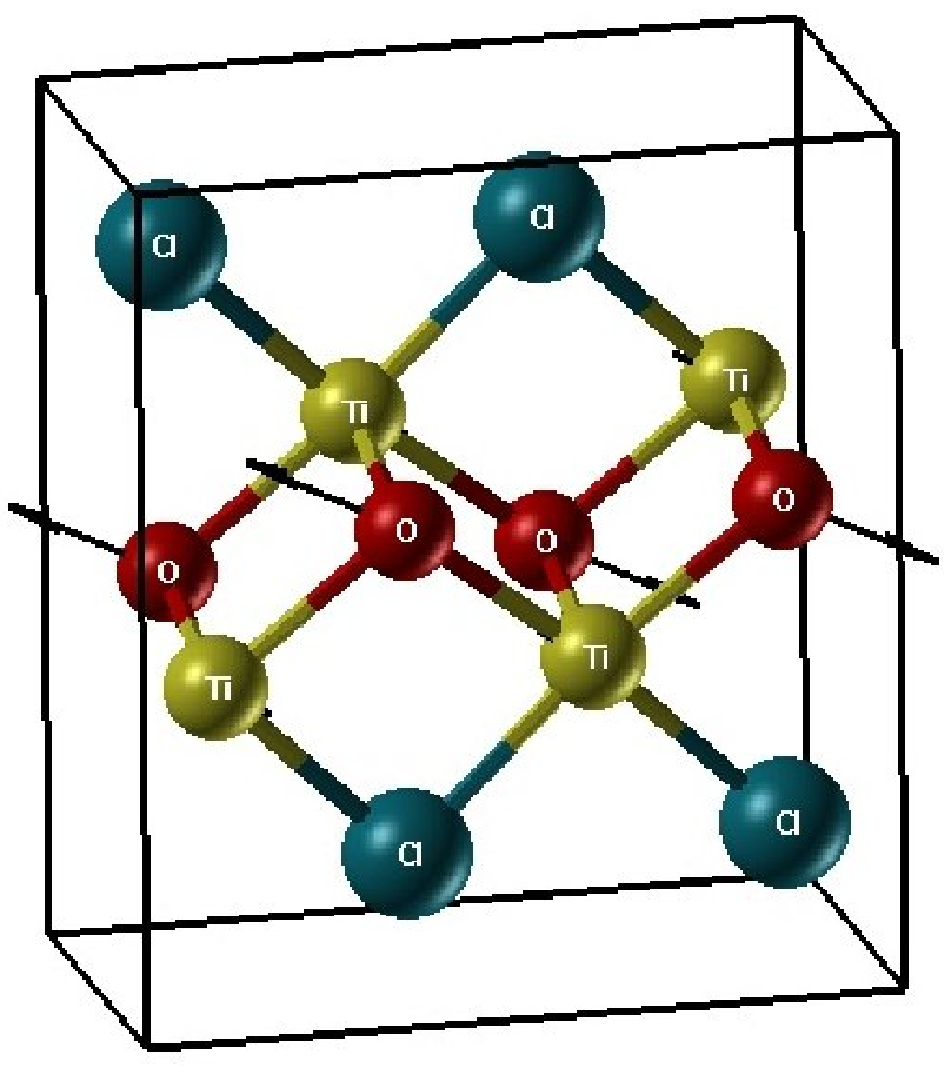}
\end{minipage}
\end{center}
\vspace{-1cm}
\begin{center}
\mbox{B$_u$}
\begin{minipage}{5.5cm}
\includegraphics[width=5.5cm]{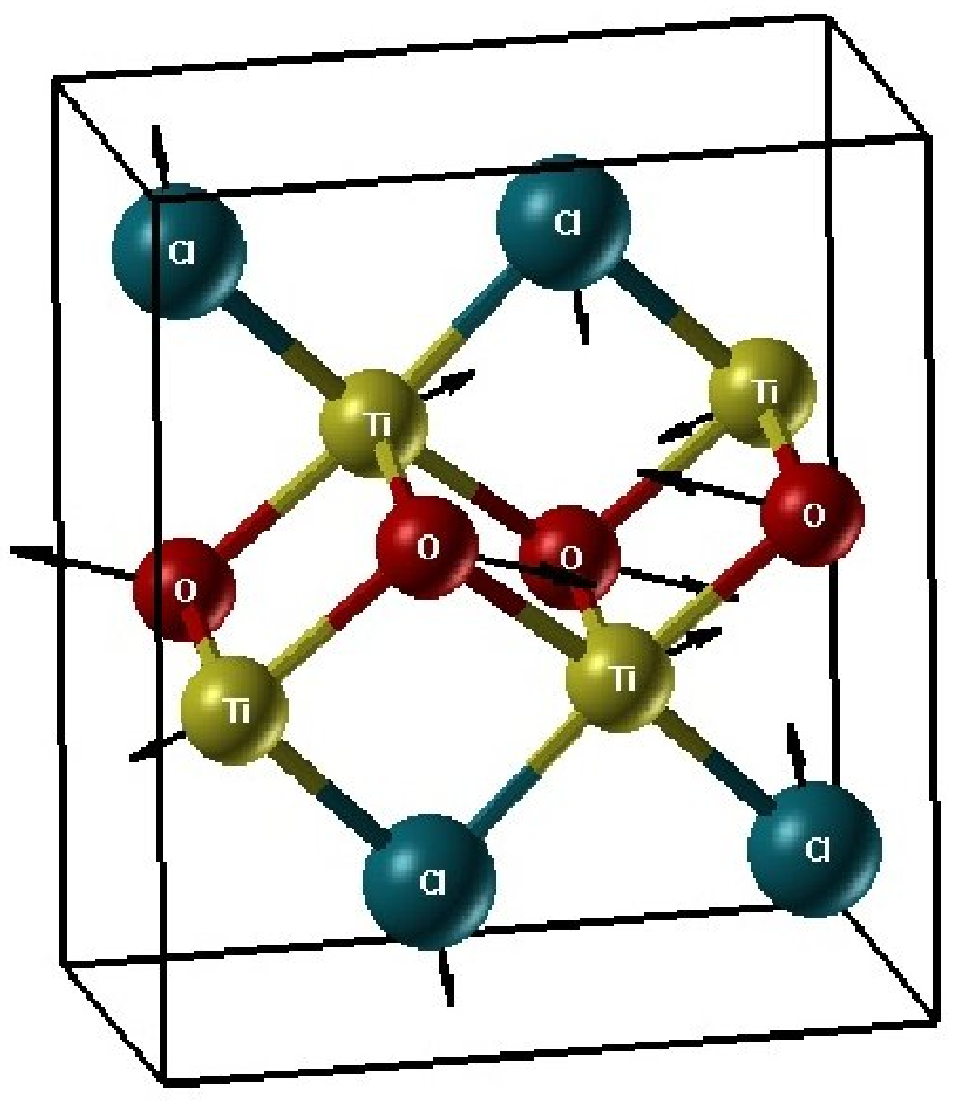}
\end{minipage}
\begin{minipage}{5.5cm}
\includegraphics[width=5.5cm]{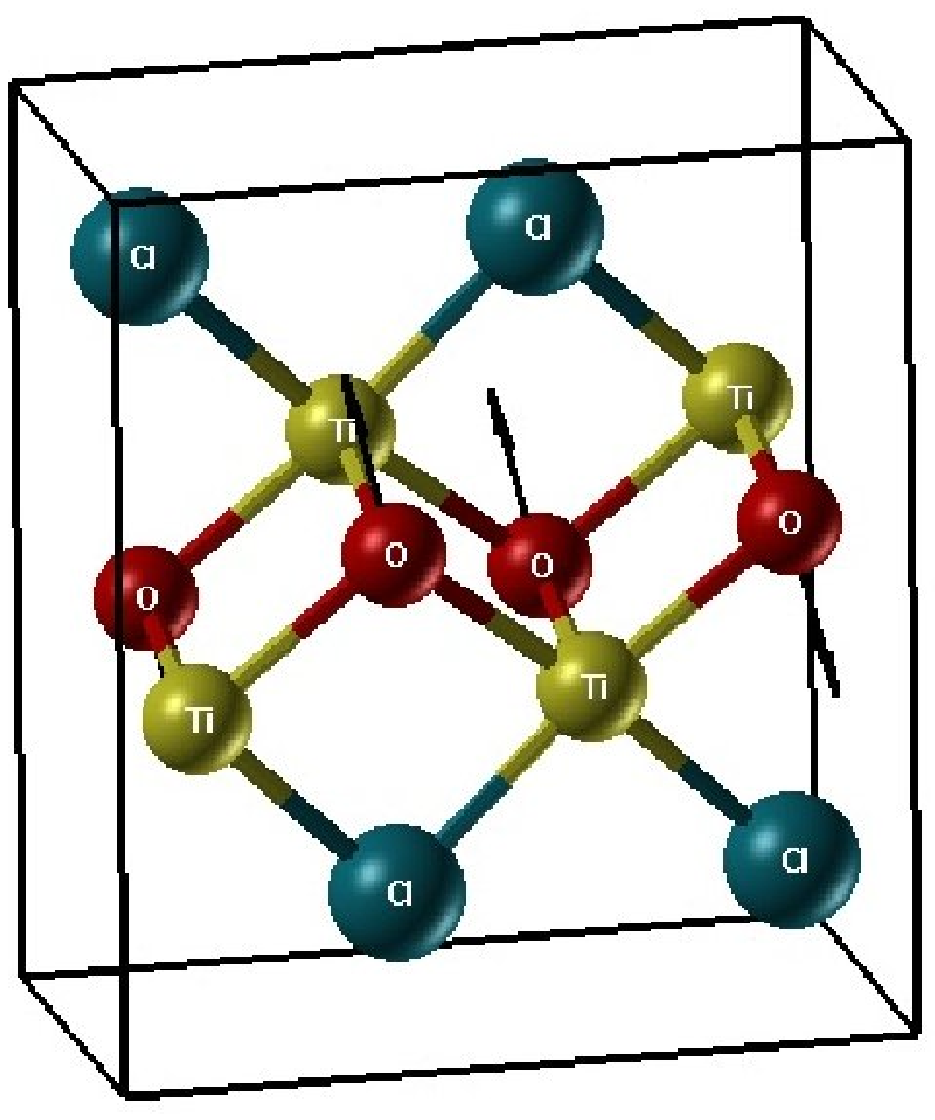}
\end{minipage}
\begin{minipage}{5.5cm}
\includegraphics[width=5.5cm]{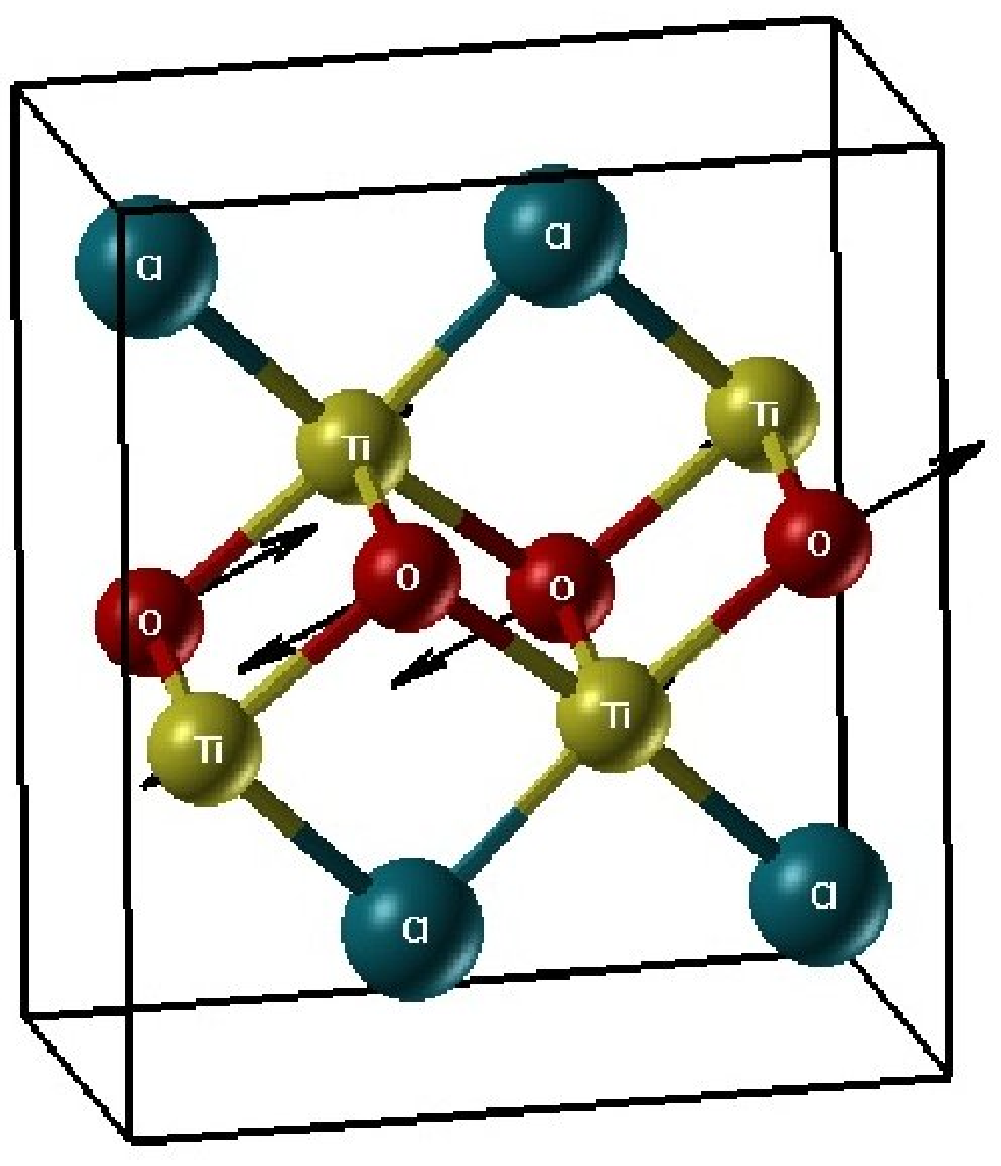}
\end{minipage}
\end{center}

\caption{(Colour online) Eigenvectors of the six A$_g$ and six B$_u$ modes originating in the low symmetry phase,
         grouped according to their degeneracy. 
         Yellow spheres are Ti atoms, red oxygen and blue chlorine.
         The system of reference is indicated for the first mode on the top-left pannel and remains 
         unchanged for the remaining modes. Length and direction of the
arrows are directly proportional to the strength and vibrational direction of the modes.}
\label{glyphs}

\end{figure*}

A peculiar feature of the two panels of Table~\ref{ltfreq} is that the strictly LS Raman
active modes are degenerate with their infrared counterparts 
i.e.
the six A$_g$ with the six B$_u$ modes  (third row in the upper
and lower panel of Table \ref{ltfreq}) and the three B$_{g}$ with the 
three A$_u$ modes (fifth row in the upper and lower panel 
of Table \ref{ltfreq}).
The six A$_g$ and six B$_u$ modes are illustrated in  Fig.~\ref{glyphs} 
in order of increasing frequency value.

\begin{figure}
\includegraphics[width=7cm]{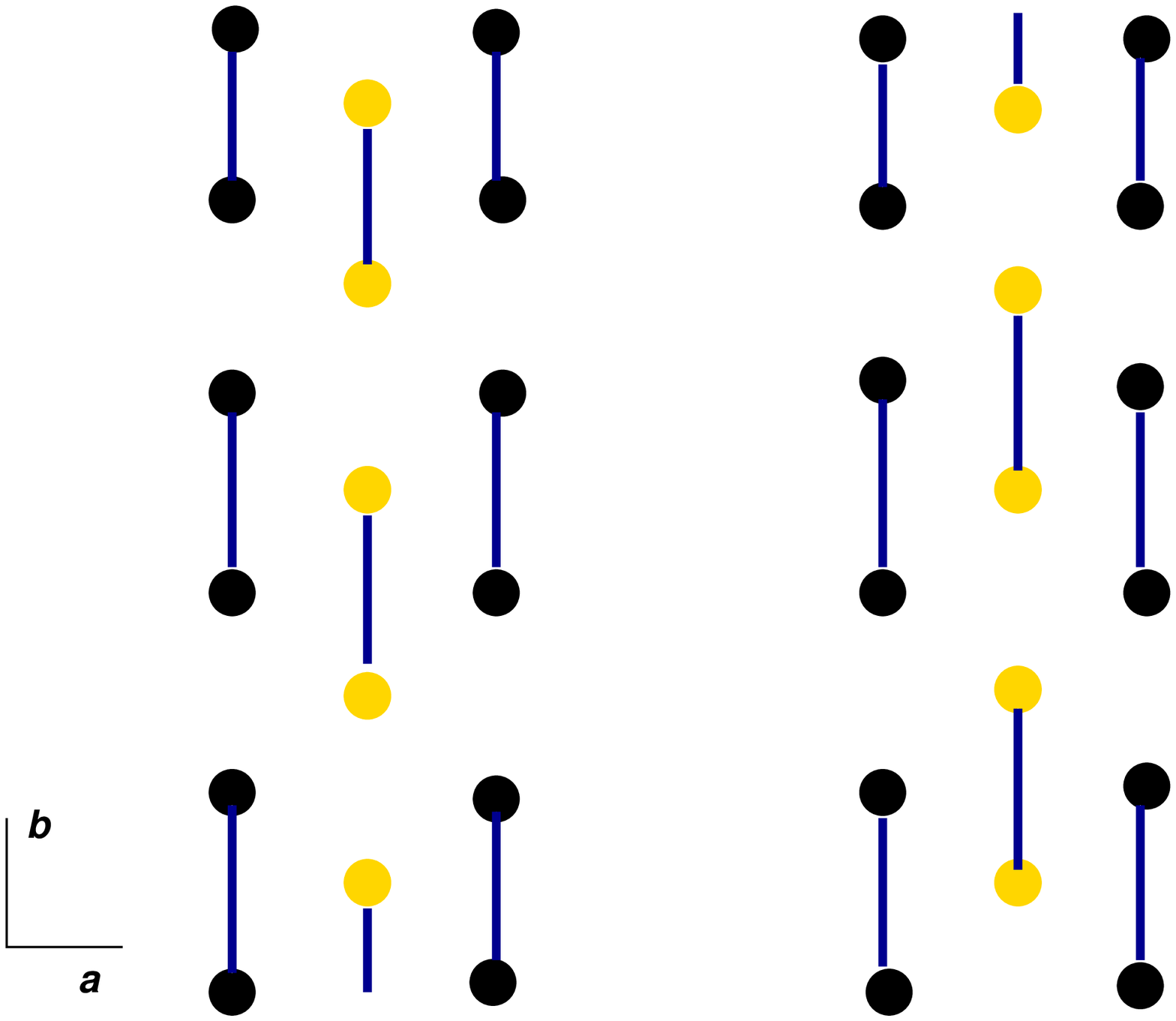}
\caption{ (Colour online) Schematic representation of the two possible
Ti-Ti bond patterns within the   $P2_1/m$ symmetry group for TiOCl.
Black and yellow symbols denote the Ti atoms on two consecutive layers of a single bilayer.
}
\label{patterns}
\end{figure}

The degenerate character of these modes is due to the perfect triangular
geometry of bonds between Ti atoms belonging to 
neighbouring chains that run along the $b$ axis. 
This geometry makes the interchain    dimerisation patterns
shown in Fig.~\ref{patterns}  
 degenerate (this feature should be observed in some zone boundary phonons of the HS phase).
In fact, if we were to consider the case of ladder-type arrangement
 of chains,  every Ti atom
would have only one nearest neighbour in the neighbouring chain, 
and the two  dimerisations shown in Fig. ~\ref{patterns}
would involve 
different lattice strains and therefore would be non degenerate.
In the present case, where one chain is rigidly shifted
 by half of the lattice parameter $b$ 
with respect to the other, every Ti acquires two 
equivalent nearest neighbours along $a$ and therefore the in-phase and out-of-phase
displacements shown in Fig.~\ref{patterns} become equivalent.
Note that this degeneracy would be split within a model
which incorporates spin-singlet fluctuations, as pointed out above,
although not significantly. 
However, in the non-dimerised phase (T $>$ T$_{c1}$) 
such a degeneracy is restored.

All the six A$_g$ and six B$_u$ modes which originate in the LS phase
represent possible dimerizations of the Ti chains within the $bc$ plane. 
The two lowest energy modes (at $175$ and at $212$ cm$^{-1}$)  keep
the Ti-O distance along $a$ (see Table~\ref{dist})
close to its optimised value, while the Ti-O distance along $b$ oscillates,
in a bond bending fashion for the mode at $175$ cm$^{-1}$ and as bond
stretching mode
for the one at $212$ cm$^{-1}$. In fact, the elastic constant associated with the Ti-O bond
along $a$ is larger than the one along $b$ due to the shorter bond length. 
The Ti-O$|_{a}$ bond sets the interchain elastic coupling and, due to its presence,
the LS modes in the $bc$ plane are not merely displacements along the $b$ and $c$ axes independently,
as one would expect if the Ti chains were not coupled.
The higher frequency modes have larger O components and do not generally preserve
the Ti-O$|_{a}$ equilibrium distance, as inferable from the length and direction of the arrows
representing the eigenvectors' components in Fig~\ref{glyphs}.

As a general comment on the six A$_g$ and six B$_u$  LS modes,
 it is interesting to
note that they show a low dimensional character.
They  have a two-dimensional configurational space ($bc$ plane) and due to the
strong interchain coupling some of them show an effective one-dimensional oscillation.
This can be seen, for example, for the modes at $175$ and at $212$ cm$^{-1}$ in Fig.~\ref{glyphs}
where the arrows on the Ti and O atoms have same length and direction,  
suggesting that the Ti-O-Ti-O plaquettes may be considered as an effective single vibrational block.
The two modes represent the bending and the stretching of the bond 
between two neighbouring plaquettes. 
These features evidence the possible role of low-dimensional structural fluctuations
as we shall discuss in subsection~\ref{strfluct}.

\subsection{Phonon assignment and Raman spectra anomalies}
\label{phass}

We note that the doubly degenerate lowest mode ($A_g$, $B_u$) at 175
cm$^{-1}$. i.e., the zone center phonon
of the LS phase and a zone boundary phonon of the HS phase,
  (Fig.~\ref{glyphs} first column, first and second raw panels)
remarkably resemble atom displacements along the spin-Peierls distortion
reported in Ref.~\onlinecite{shaz:100405}
and Ref.~\onlinecite{schonleber} (compare Fig. 1(b) 
of Ref.~\onlinecite{shaz:100405} and Fig. 3 of
Ref.~\onlinecite{schonleber}  with Fig.~\ref{glyphs} first column, second raw
panel).
The slight discrepancy in the eigenvectors' directions points towards an overestimation,
in our approach of the elastic constant stiffness of Ti-Ti bond along
$b$, which is consistent with the b length overestimation in section~\ref{vol}.

  This frequency at 175 cm$^{-1}$ is close to the frequency of an 
anomalous mode observed 
in the Raman spectrum of Ref.~\onlinecite{Lemmens_04} (named $\alpha$)
which consists of a broad peak centered around 160 cm$^{-1}$ at room temperature.
As the temperature decreases, the peak position shifts towards lower frequency values
and its broadening reduces until it becomes indistiguishible
from the Raman A$_g$ mode allowed in the low symmetry phase.
Due to its Lorentzian lineshape, this feature is attributed to 
low dimensional energy fluctuations of the spin system which manifest themselves in the form
of quasielastic scattering of the Raman light and decrease as the temperature
approaches the spin-Peierls transition point~\cite{Lemmens_04}.
It is worth pointing out that this broad feature is detected 
when the Raman light is  polarised along the $b$-axis 
(along which the spin chains lie) while it is 
totally absent when the light is polarised along the 
$a$-axis~\cite{Lemmens_04}.

Recently, the evolution with temperature of the Raman active 
A$_g$ modes has been studied by Fausti {\em et al.}~\cite{fausti}.
At low temperature, along with the three unchanged HS $A_g$ phonons, 
they find nine additional modes as required by group theory analysis
(see Table~\ref{ltfreq} upper pannel). 
Four of these nine modes persist above the spin-Peierls transition  and 
become broader losing in intensity as the temperature increases. 
These modes are also visible in the Raman spectrum of 
Ref.~\onlinecite{Lemmens_04} at low temperatures
and the lowest frequency mode coincides with the feature $\alpha$ described above.  

Combining the information of the Raman spectrum of Ref.~\onlinecite{fausti} (where 
not only light polarisation $bb$ was used but also $bc$) 
with the information contained in Table ~\ref{ltfreq}
and in the phonon displacement eigenvectors (not shown), 
we are able to assign the experimental Raman frequencies to the calculated 
modes (Table~\ref{assign}). 
In the following reasoning we assume that the possible splitting of the LS modes' degeneracy
due to quantum spin effects does not alter the picture qualitatively.
We have already pointed out the fact that the HS B$_{3g}$ modes remain almost identical
to their  LS  (LS $A_g$) counterpart, as shown in Tables~\ref{HTph} and ~\ref{ltfreq}. 
At high temperature, the B$_{3g}$ modes are active with light polarisation $bc$ 
and therefore we expect their identical LS counterparts to have a similar activity. 
In the spectrum of Ref.~\onlinecite{fausti}, the two modes at 178.5 and 524.3 cm$^{-1}$ 
are predominantly active with light polarisation $bc$, but very weakly active 
with polarisation $bb$ (this is confirmed in the spectrum of Ref.~\onlinecite{Lemmens_04}). 
We therefore assign these two modes to the two LS A$_g$ modes
with displacement along $b$ at 140 and 431  cm$^{-1}$.
In order to assign the third mode of the same symmetry, we note 
that the modes at 296.5 and 305.3  cm$^{-1}$ in the
spectrum (with light polarisation $bb$) of Ref.~\onlinecite{Lemmens_04}
are weakly active. Assuming a systematic underestimation of the frequencies 
due to  shortcomings of the theory (which will be discussed below) 
we assign the mode at 305.3  cm$^{-1}$ to the calculated value at 289  cm$^{-1}$.
The remaining six LS A$_g$ modes are then assigned straightforwardly in order of increasing value
(see Table~\ref{assign} for their assignment) and the three HS A$_g$ modes remain unaltered at low temperature 
(compare Table~\ref{assign} with Table~\ref{HTph} for their assignment).

\begin{table}
\caption{Mode assignment for the twelve Raman active A$_g$ modes listed in Table~\ref{ltfreq}. 
L.S. labels phonons that disappear at the SP transition T$_{c_1}$, 
I.P. those that survive up to the intermediate phase transition T$_{c_2}$, 
H.S. represents the high symmetry phase phonons. 
}
 
\begin{tabular}{cccccc}
           \multicolumn{6}{c}{L.S.}\\   
\hline
\hline
  exper. & 145.8 & 178.5 & 211.5 & 305.3 & 524.3 \\
  theor. & 212   & 140   & 285   & 289   & 431   \\
\hline
\end{tabular}

\vspace{.3cm}

\begin{tabular}{ccccc}
           \multicolumn{5}{c}{I.P.}\\
\hline
\hline
  exper. & 131.5 & 296.5 & 322.6 & 387.5  \\
  theor. & 175   & 323   & 382   & 490    \\
\hline
\end{tabular}

\vspace{.3cm}

\begin{tabular}{cccc}
           \multicolumn{4}{c}{H.S.}\\
\hline
\hline
  exper. & 203.5   & 365.1    & 431    \\ 
  theor. & 203     & 342      & 443    \\
\hline
\end{tabular}

\label{assign}

\end{table}

By direct comparison, we find that while the three HS A$_g$ modes are in good agreement
with experiment, the three LS A$_g$ modes along $b$ (B$_{3g}$ modes in the HS phase) are underestimated 
by 10-20\% and the remaining six A$_g$ modes of the LS phase are overestimated 
by 20-30\%. Possible origin of the above discrepancies could be:
1) the lack  of a screened exchange term in the Hamiltonian which also
   causes an overestimation of $b$;
2) the lack of spin-singlet dynamical correlations in the B3LYP Hamiltonian which usually
renormalise the phononic spectrum and, as mentioned in section IV,
   may be responsible for the overestimation of the lattice parameter $b$.

Based on the above assignment, we are able to conclude that the  mode at 175 cm$^{-1}$
most probably evolves into the high temperature $\alpha$ feature of Ref.~\onlinecite{Lemmens_04}.
The remaining five LS A$_g$ modes show also dimerisation patterns which may be tentatively 
related to the broadening of their Raman spectra~\cite{Lemmens_04,fausti} due to  
spin fluctuations.
Note, however, that two of the modes disappear at T$_{c_1}$ due to a symmetry change in the intermediate phase.
It would be interesting to perform a phonon calculation within the 
space group 13 ($P2/c$), suggested in Ref.~\onlinecite{fausti} as the intermediate
phase symmetry, to countercheck the validity of our assignment as well as of the suggested
space group.

\subsection{Role of pretransitional fluctuations and the spin-Peierls instability}
\label{strfluct}

 As seen above, the computed LS phonon calculations enable i) the 
  comparison with Raman and infrared spectroscopy
measurements performed for the zone center modes in 
 the spin-Peierls phase as we showed in the previous section and
 ii) the prediction of the behavior of some zone boundary
phonons in the HS phase as we will discuss in the following.

From our calculations,
the $\alpha$ mode can be interpreted as a precursive mode characterised 
by a strong spin-Peierls activity.
In fact, this has a very broad lineshape and shows a large softening
in a temperature range between T$_{c1}$ and room temperature~\cite{Lemmens_04}.
These features are associated with energy fluctuations of the magnetic and structural system.
Concerning the latter, above the transition temperature T$_{c_2}$ strong incommensurate 
fluctuations have been detected in X-ray diffraction experiments~\cite{schonleber}
and they are seen to persist up to room temperature.
Our phonon calculation supports the lattice fluctuation scenario with effectively one-dimensional
atom displacement paths, as shown in subsection~\ref{phass} for the lowest frequency modes. 
In addition, anomalies in the electronic structure are also found  in the ARPES spectra 
of the high temperature phase~\cite{Hoinkis_05} which have been given little explanation so far.

Recently, the authors of Ref.~\onlinecite{Abel07}  reported the observation of
 a  zone boundary acoustic phonon in TiOCl which
 softens at T$_{c_2}$\cite{Gros98,Dobry07}. They identify this mode as the one that drives the
spin-Peierls transition and estimate its bare phonon frequency to be $\Omega_0$ $\sim$ 217 $cm^{-1}$.
In our calculation of the relevant zone boundary modes in the HS for the unit
 cell doubling along $b$ (section VI A)
we found that the doubly degenerate mode ($A_g$, $B_u$) at 175 cm$^{-1}$ shows
features of a spin-Peierls phonon in its vibrational pattern
 (see Fig.~\ref{glyphs}).  While we cannot say at this point whether this corresponds
 to an acoustic or to an optical zone boundary phonon of the HS phase, its
 energy is close to the bare phonon energy estimated by Abel {\it et al.}\cite{Abel07}.
Moreover, one could then speculate that there is a
direct relation between the $\alpha$ phonon observed
in Raman spectroscopy 
with the interpreted spin-Peierls phonon observed in inelastic X-ray experiments.

A last remark concerns the incommensurate region between $T_{c_2}$ and $T_{c_1}$.  This region shows incommensurate
 wave vectors shifted from the commensurate wave vector $(0,\frac{1}{2},0)$  by 
the amount $(\pm \Delta H, \pm \Delta K,0)$. It has been argued that the origin
of this incommensurability originates from competing arrangements of bond patterns
in TiOCl\cite{Rueckamp_05,schonleber,Clancy_07} and more recently in terms of a model
of discommensurations separating commensurate from dimerized regions\cite{Abel07}.
The existence of degenerate modes enforces the argument of competing patterns in the
incommensurate region with possible coexisting domains.

\section{Conclusion}

In summary, we have presented  within hybrid density functional theory  
 a complete description of the phonon modes (Raman and infrared active)
in TiOCl both  in the space group $Pmmn$ (high symmetry phase) and
 in the space group $P2_1/m$  which is the symmetry of the spin-Peierls phase. 
Comparison with experiment and with results obtained from
other functionals shows that   non-local 
correlations as implemented in  the hybrid B3LYP approach are important
in order to obtain an improved description of the modes.
This study supports the correlated nature of TiOCl.

Our analysis of the zone center high-symmetry modes (space group $Pmmn$)
 and  some relevant zone boundary high-symmetry modes (space group $P2_1/m$)
allows us to identify the anomalous $\alpha$ mode
observed in Raman scattering as a probable precursor of the spin-Peierls mode
strongly coupled to the spin system.  Moreover, the 
triangular geometry of bonds between Ti atoms belonging to 
neighboring chains running along $b$ leads to the existence
of six A$_g$ zone boundary modes degenerate with six B$_u$ zone boundary modes 
(for the undistorted system), a phenomenon which has not been observed in other
known spin-Peierls systems and which contributes to the anomalous
behavior of TiOCl.

Some of the low symmetry modes show effectively one-dimensional
atomic displacement paths supporting the important role of lattice fluctuations
in addition to the magnetic ones.

\section*{Acknowledgments}

We would like to thank R. Claessen, C. Gros, M. Gr\"uninger, H.O. Jeschke, P. Lemmens, T. Saha-Dasgupta,
M. Sing , A. Dobry, M. Mostovoy, D. Khomskii and G. Mallia for useful discussions.  
R.V. thanks the Deutsche Forschungsgemeinschaft
for financial support through FOR412 and SFB/TRR49 grants and the NSF  through Grant. No. PHY05-51164.

\bibliography{paper.bib}
\bibstyle{abbrev}
\nocite{*}

\end{document}